\documentclass[final,5p,times,twocolumn]{elsarticle}

\usepackage{framed,multirow}

\usepackage{amssymb}
\usepackage{amsmath}

\usepackage{subcaption}
\usepackage{comment}
\usepackage{xspace}
\usepackage{amsmath}
\usepackage{amssymb}

\usepackage{booktabs}
\usepackage{hyperref}

\usepackage{enumitem}

\newcommand{\diffshadow}{\textit{DiffusionShadow}\xspace}

\newcommand{\eg}{\MakeLowercase{e.g.,}\xspace}

\newcommand{\etal}{\MakeLowercase{et al.}\xspace}

\newcommand{\dataname}[1]{{\texttt{#1}}}

\usepackage{xcolor}
\usepackage[normalem]{ulem}

\newcommand{\rev}[1]{#1}

\makeatletter
\def\ps@pprintTitle{%
  \let\@oddhead\@empty
  \let\@evenhead\@empty
  \let\@oddfoot\@empty
  \let\@evenfoot\@empty
}
\makeatother

\begin{document}

\begin{frontmatter}



\title{DiffusionShadow: Diffusion-based Shadow Caching for Neural Volume Rendering}

\author[aff1]{Kai-Chen Tung}
\ead{kctung@ucdavis.edu}
\author[aff2]{Qi Wu}
\ead{qiwu@nvidia.com}
\author[aff1]{David Bauer}
\ead{davbauer@ucdavis.edu}
\author[aff3]{Mengjiao Han}
\ead{hanm@anl.gov}
\author[aff3]{Silvio Rizzi}
\ead{srizzi@anl.gov}
\author[aff1]{Kwan-Liu Ma}
\ead{klma@ucdavis.edu}

\address[aff1]{University of California, Davis, Davis, CA, United States}
\address[aff2]{NVIDIA Spatial Intelligence Lab, Santa Clara, CA, United States}
\address[aff3]{Argonne National Laboratory, Lemont, IL, United States}




\begin{abstract}
Implicit neural representations (INRs) have gained momentum in scientific visualization due to their compactness and scalability to large datasets, making them well suited for integration with direct volume rendering (DVR). However, real-time volume rendering of INR with advanced illumination effects, such as shadows, remains computationally expensive, as evaluating shadow terms via ray marching is costly. Alternatively, precomputing and storing shadows for many lighting directions is prohibitive in both memory and storage. To address this, we introduce a diffusion-based shadow caching framework that compresses a vast set of pre-calculated shadow INRs into a single diffusion model. Rather than focusing on generalizing to unseen directions, our method effectively memorizes and reconstructs a dense set of pre-trained lighting conditions on the fly. We first encode a collection of shadow coefficient volumes as shadow INRs, and then train a diffusion model conditioned on lighting direction to predict the corresponding shadow INR weights at inference time. This design integrates directly with standard INR renderers without additional runtime sampling. Experiments show that our approach achieves faster rendering than traditional methods while bypassing the massive storage bloat of independent INRs, producing shadows that closely match most of the reference results.
\end{abstract}




\end{frontmatter}


\section{Introduction}
Implicit neural representations (INRs) have recently shown strong potential for scientific visualization due to their capability to approximate arbitrary volumetric fields in an extremely compact form while preserving high-fidelity reconstruction quality~\cite{wu2024distributed, lu2021compressive}. By nature, these representations support continuous point queries in the volume domain, which makes them suitable for traditional direct volume rendering (DVR) algorithms, as samples can be queried directly without decoding an explicit representation of the volume~\cite{wu2023interactive}.

However, enabling shadow-based visualization for directly rendering these volumetric INR representations presents significant challenges. 

Traditional methods, such as casting secondary shadow rays, do not suit this task well. First, tracing divergent secondary rays disrupts the wavefront-style rendering schedule required to efficiently batch network queries. Second, computing a shadow ray for every sample introduces a prohibitive $\mathcal{O}(N^2)$ computational complexity (assuming $N$ samples per primary ray). 
This overhead is further amplified when the volume is represented solely as an INR instead of a raw voxel grid, since each neural network query is orders of magnitude more expensive than a direct voxel lookup. Together, these issues severely undermine the feasibility of interactive visualization with advanced illumination for INRs.

Although several works have utilized the representational power of INRs to approximate and cache costly illumination terms in the rendering equation---such as shadow fields~\cite{wu2023hyperinr}, photon-mapping fields~\cite{bauer2023photon}, and irradiance fields~\cite{muller2021real,mildenhall2020nerf}---
these pipelines still require training and storing a separate INR for each specific lighting condition. In a typical scenario where a user might adjust the light direction freely, precomputing and storing thousands of such INRs would incur prohibitive memory and storage costs, limiting practical deployment.

On the other hand, denoising diffusion models~\cite{ho2020denoising} have recently emerged as the state-of-the-art method for high-fidelity synthesis. By formulating synthesis as learning the reverse of a noise-adding Markov process, diffusion methods avoid the training instabilities common in adversarial approaches~\cite{goodfellow2014generative}. 

The same paradigm has been extended to 3D~\cite{poole2022dreamfusion,lorraine2023att3d,lin2023magic3d,zeng2022lion,ren2024xcube} and even enables the creation of content parameterized as INRs~\cite{erkocc2023hyperdiffusion, wu2024blockfusion, wang2023rodin}. 
Because diffusion models can be conditioned on external signals~\cite{nichol2021glide}, they are well suited to address the aforementioned storage limitations. Instead of keeping a vast library of individual shadow INRs on disk, we can train a conditional diffusion model to memorize and reconstruct the weights of these INRs. Given a target lighting condition from the pre-trained set, the model effectively recovers the corresponding INR weights on the fly. 

In this work, our main driving application is enabling shadow-based visualization for directly rendering volumetric INR representations. To achieve this, we develop \diffshadow, a diffusion-based shadow caching framework that predicts shadow INR weights given specific lighting directions. We hypothesize that diffusion models can compress the large, discrete distribution of shadow coefficients across lighting directions into a unified weight space. By using the generative model to directly predict a separate SIREN-based shadow INR, we can co-render both the data INR and the shadow field, 
\rev{replacing expensive secondary-ray queries to the data INR with a single shadow INR evaluation per primary sample, thereby accelerating rendering.}

Our pipeline first trains a collection of SIREN INRs~\cite{sitzmann2020implicit} on shadow coefficients computed under a dense set of sampled lighting directions. We then train a diffusion model, conditioned on the lighting-direction vector, to synthesize the shadow INR weights representing the corresponding shadow field. Once trained, the reconstructed shadow INRs, together with the data INR, integrate directly into interactive volume rendering systems. This supports real-time exploration with realistic shadows under the pre-trained lighting directions without incurring the high computational cost of on-the-fly shadow evaluation or the massive storage bloat of independent neural networks. \rev{We evaluate shadow reconstruction quality, visual fidelity, and rendering performance, demonstrating that our shadow INR achieves at least a 20$\times$ rendering speedup across all test datasets over naïve INR-based shadow ray rendering while maintaining realistic shadows.} 
We further compare our method with existing approaches and conduct ablation studies to analyze the impact of key design choices.

We summarize our contributions as follows:
\begin{itemize}[itemsep=2pt, topsep=2pt, parsep=0pt]
  \item \textbf{Diffusion-based Shadow INR Prediction}: 
  We introduce a novel diffusion caching framework that enables direct shadow INR prediction conditioned on lighting directions, eliminating the need for per-lighting-direction optimization at inference time.
    
  \rev{\item \textbf{Efficient Shadow Rendering for INR-based Volume Visualization}: The generated shadow INRs are seamlessly integrated into an interactive INR-based volume renderer and co-rendered with the data INR, eliminating costly secondary-ray evaluation and enabling real-time rendering with realistic shadows.}
  
  \item \textbf{Comprehensive Evaluation}: We conduct extensive comparisons and ablation studies to analyze reconstruction quality, rendering performance, and the impact of key design choices, including geometry loss, rendering loss, and lighting direction sampling.
\end{itemize}
\section{Related Work}
We build on recent progress in the domain of INRs, INR parameterization, and machine learning-based global illumination. We also take inspiration from diffusion models, which have shown strong capabilities for synthesizing high-quality 3D content. Below, we survey the most relevant developments in each of these areas.

\subsection{Implicit Neural Representations (INRs)}
INRs~\cite{xie2022neural} have gained widespread attention following the success of Neural Radiance Fields (NeRF)~\cite{mildenhall2020nerf}, which demonstrated that continuous coordinate-based networks can effectively model complex 3D scenes. While NeRF primarily focuses on novel view synthesis and scene reconstruction in computer vision, recent works have extended the concept of INR to the scientific visualization domain, where they show great potential for compressing and rendering large-scale scientific data due to their compact size and inherent support for random access reads. NeurComp~\cite{lu2021compressive} is one of the early works to use INRs to compress volumetric scalar field data. Coordnet~\cite{han2022coordnet} later introduced a unified coordinate-based INR framework to generalize on various tasks such as spatial and temporal super resolution. In addition to compression purpose, INR can also facilitate interactive visualization. Wu~\etal~\cite{wu2023interactive} utilized INRs with a multi-resolution hash encoding~\cite{muller2022instant} along with a sample streaming rendering algorithm and out-of-core sampling to achieve interactive visualization even for terascale datasets.

Positional encoding layers are also essential components of INR models, transforming input coordinates into feature spaces that facilitate compact and efficient learning. Fourier feature mapping~\cite{tancik2020fourier} and the positional encoding used in NeRF~\cite{mildenhall2020nerf} employ sinusoidal transformations to capture high-frequency variations, while Instant-NGP~\cite{wu2023interactive,muller2022instant} introduces multi-resolution hash encoding to reduce network size requirements without sacrificing fidelity. Triplane encoding~\cite{chen2022tensorf, fridovich2023k} is another popular type of encoding, which decomposes 3D scenes into three orthogonal 2D feature planes. One important benefit of triplane encoding is that the trained triplanes can be directly used as inputs to 2D GANs~\cite{chan2022efficient} or image-based diffusion models~\cite{wu2024blockfusion} for generative modeling. Following this idea, we represent shadow coefficient volumes using a triplane encoding INR to exploit the efficiency of 2D diffusion models for learning, reconstruction and extrapolation.

\subsection{INR Parameterization}

Another line of research explores generating implicit neural representations (INRs) through weight-space prediction. In this process a network produces the parameters of another network. Ha~\etal~\cite{ha2016hypernetworks} first introduced this concept with hypernetworks, which generate the weights of a target network, allowing the target network to adapt its parameters dynamically based on input conditions. Building on this idea, Klocek~\etal~\cite{klocek2019hypernetwork} applied hypernetworks to produce INR weights for image super-resolution, while Sitzmann~\etal~\cite{sitzmann2019scene} proposed Scene Representation Networks (SRNs), where an MLP hypernetwork is used to generate the weights of a neural network that represents a specific 3D scene. More recently, HyperINR~\cite{wu2023hyperinr} extended the concept to high-dimensional scientific data by interpolating INR encoding weights for time-varying volumes. Similarly, in the scientific visualization domain, HyperFLINT~\cite{gadirov2025hyperflint} employs a hypernetwork to model ensemble simulations, enabling flow field estimation and temporal interpolation conditioned on simulation parameters. Besides, HyperDiffusion~\cite{erkocc2023hyperdiffusion} demonstrated that diffusion models can directly learn the distribution of INR weight space to synthesize 3D shapes and 4D animations. In this spirit, our work focuses on INR weight generation, where a pretrained diffusion model learns to synthesize the triplane encoding weights representing shadow coefficient volumes.

\subsection{Machine Learning-Based Global Illumination}
Global illumination (GI) encompasses a broad range of light transport effects, including ambient occlusion, indirect lighting, and shadows. Achieving real-time rendering with high-quality global illumination remains computationally expensive, motivating the use of machine learning models to approximate these effects efficiently.
A major line of research focuses on screen-space methods that operate directly on rendered images or G-buffers to approximate GI. Gao~\etal~\cite{gao2022neural} trained neural networks using shading attributes, view directions, and light directions to predict indirect illumination, which can be combined with direct illumination for realistic rendering. Nalbach~\etal~\cite{nalbach2017deep} employed convolutional neural networks (CNNs) to synthesize shading effects such as ambient occlusion from deferred shading buffers. Similarly, Deep Illumination~\cite{thomas2017deep} utilized generative adversarial networks (GANs) to learn mappings from G-buffers and direct illumination buffers to full global illumination images.
Beyond screen-space rendering, several works extend deep learning to volumetric domains for achieving GI effects directly within 3D data. DVAO~\cite{engel2020deep} applied 3D CNNs to predict ambient occlusion volumes and explored different strategies for conditioning the network on transfer functions. Bauer~\etal~\cite{bauer2023photon} represented photon fields as neural networks and employed a neural path tracer to enable interactive global illumination in volumetric visualization. Unlike approaches that generate global illumination in the 2D image domain, operating directly within the volumetric domain enables integration into interactive visualization pipelines. Our method utilizes precomputed shadow coefficient volumes and employs diffusion-based prediction to reproduce shadow effects under direct volume rendering. The predicted results can be immediately incorporated into the rendering process without additional postprocessing.

\subsection{Denoising Diffusion Models}
Recent years have seen denoising diffusion models~\cite{ho2020denoising, sohl2015deep} become widely adopted in generative modeling. Denoising Diffusion Probabilistic Models (DDPM)~\cite{ho2020denoising} first demonstrated that diffusion models can synthesize high-quality images, rivaling other generative models such as GANs. Subsequent works~\cite{dhariwal2021diffusion, nichol2021improved} further improved their architectures and training strategies, achieving higher sample quality, greater efficiency, and better coverage of the data distribution. However, training diffusion models directly in pixel space remains computationally expensive. To address this, latent diffusion model (LDM)~\cite{rombach2022high} proposed learning diffusion in a compact latent space, significantly reducing computational cost while preserving high-fidelity image synthesis. More recent studies have also employed diffusion models to explore and overcome the challenges of generating images with complex illumination effects, such as shadows~\cite{sarkar2024shadows}, reflections~\cite{dhiman2025reflecting}, and caustics~\cite{10.2312:egs.20251030}.

Beyond image-space generation, diffusion models have also been extended to 3D domains. For example, Zheng~\etal~\cite{zheng2023locally} applied a 3D U-Net-based diffusion model directly to signed distance fields for 3D shape synthesis. SSDNeRF~\cite{chen2023single} used diffusion models in conjunction with NeRFs to achieve 3D reconstruction from images. Wang~\etal~\cite{wang2023rodin} represented 3D avatars using triplanes and performed diffusion in the triplane space, while Shue~\etal~\cite{shue20233d} adopted a similar triplane-based diffusion approach for mesh data. To further enhance scalability and generative quality, BlockFusion~\cite{wu2024blockfusion} introduced using an autoencoder to compress triplanes into a latent space, making them more suitable for diffusion models to learn and enabling large-scale 3D generation.

\section{Methodology}
In this section, we first describe data preparation and the training pipeline used to generate shadow INR weights. Our training pipeline adopts a two-stage procedure inspired by HyperDiffusion~\cite{erkocc2023hyperdiffusion}: (1) SIREN network overfitting (we use "overfitting" in the INR sense of fitting a network to a single signal), and (2) diffusion model training. Each stage is detailed in a dedicated subsection. We then explain the conditioning strategy and loss functions for the diffusion model and conclude with network design considerations.

\begin{figure*}[htbp]
    \centering
   \includegraphics[width=1.0\textwidth]{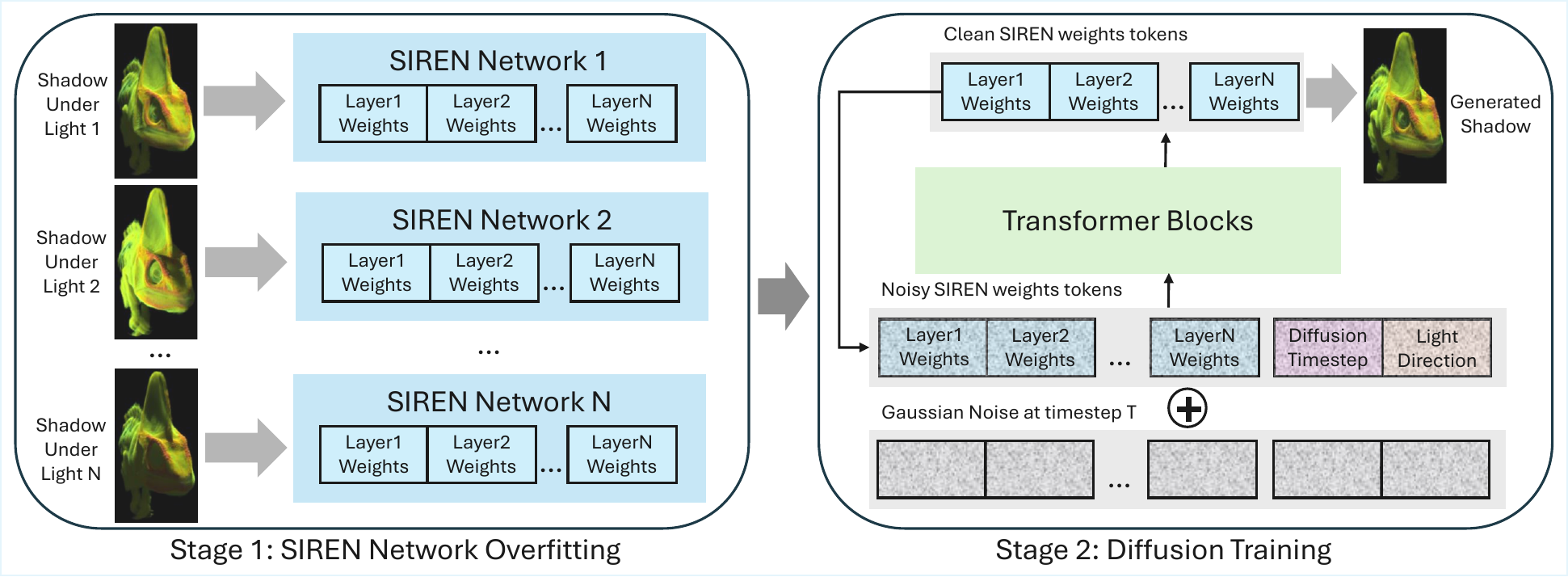}
    \caption{Overview of our two-stage training pipeline. (1) SIREN Network Overfitting: encode each shadow coefficient volume as a SIREN network. (2) Diffusion Training: synthesize SIREN network weights conditioned on lighting directions to generate shadows under pre-trained lighting directions.}
    \label{fig:overall_framework}
\end{figure*}

\subsection{INR of Shadow Coefficients}

The single-scattering model~\cite{max2002optical} is an approximate yet widely used method for volume rendering to achieve realism. It is more efficient than explicitly simulating multiple-scattering effects while still capturing salient features of volumetric data. This model augments the standard emission-absorption volume-rendering equation with a reflection term of external illumination:
\begin{equation}\label{eq:single-scattering-model}
\begin{split}
    L(\boldsymbol{x_0}, \boldsymbol{\omega})
    = \int T_{\boldsymbol{\omega}}(t, 0)~r(\boldsymbol{x_t}, \boldsymbol{\omega}, \boldsymbol{\omega}') g(\boldsymbol{x_t}, \boldsymbol{\omega}') L_i(x_t, \boldsymbol{\omega})\, dt,
\end{split}
\end{equation}
where $r$ is the BRDF, $L_i$ is the incoming illumination from an infinitely distant directional light source $\boldsymbol{\omega}'$, and $g$ is the shadow-coefficient term representing the transmittance of the medium along the path from the point $\boldsymbol{x}_t$ to the light source.

A na\"ive approach to computing shadow effects in direct volume rendering casts a secondary shadow ray toward the light source at every ray-marching sample. At each sample point, the shadow coefficient is the transmittance from the sample position $\boldsymbol{x}_t$ to the boundary of the data volume along the light direction $\boldsymbol{\omega}':$
\begin{equation}\label{eq:shadow-coefficient}
\begin{split}
g(\boldsymbol{x_t}, \boldsymbol{\omega}') = e^{- \int_0^{\infty} \mu(x_t - t \boldsymbol{\omega}')\, dt},
\end{split}
\end{equation}
where $\mu$ is the extinction coefficient (assuming $\mu=0$ outside the volume, and $\boldsymbol{\omega}'$ points toward the light source). However, for directly rendering volumetric INR representations, evaluating this integral via secondary ray marching is computationally prohibitive. It imposes an $\mathcal{O}(N^2)$ computational complexity---where $N$ is the average number of samples along a primary ray---and disrupts the wavefront execution necessary for efficient neural network querying. Alternatively, rasterizing precomputed shadow coefficients into a dense grid significantly increases memory usage, undermining the compression benefits of the data INR.

To address these issues, we represent shadow coefficients as implicit neural representations (INRs), defining a conditional function $g^{\boldsymbol{\omega}'}$ that maps a spatial location $(x,y,z)$ to a scalar value:
\begin{equation}\label{eq:shadow-coefficient-inr}
g^{\boldsymbol{\omega}'}: \mathbb{R}^3 \rightarrow \mathbb{R},~(x,y,z) \mapsto g^{\boldsymbol{\omega}'}(x,y,z) = v.
\end{equation}

To facilitate downstream diffusion training, we use SIREN networks~\cite{sitzmann2020implicit}, composed of fully-connected layers with sine activation functions, to model each shadow-coefficient function under various lighting directions $\boldsymbol{\omega}'$, as illustrated in~\autoref{fig:overall_framework}.

Unlike piecewise-linear ReLU activations, the infinitely differentiable sine activations in SIREN networks are well-suited for modeling complex high-frequency signals and achieve fast convergence with proper initialization~\cite{sitzmann2020implicit}.

For a given light direction, we train a corresponding SIREN network using a shadow-coefficient sampler that computes coefficients directly from the raw volume at arbitrary spatial coordinates via ray marching. This eliminates the need to store dense precomputed coefficient grids per lighting direction. 
However, storing independent INRs for thousands of lighting directions still introduces substantial storage bloat. To overcome this limitation, we rely on the high-capacity memory of diffusion models: we train a single diffusion model to compress a large, dense set of INRs representing shadow volumes under uniformly sampled lighting directions. Instead of focusing on generalizing shadows for unseen lighting conditions, the model reconstructs the appropriate pre-trained shadow INR weights on the fly, effectively caching thousands of INRs in one unified parameter space. By co-rendering the base data INR alongside this reconstructed shadow INR, we evaluate both fields concurrently. This reduces the runtime complexity from $\mathcal{O}(N^2)$ to $\mathcal{O}(2N)$ and maintains the wavefront execution required for efficient neural rendering.

\subsubsection*{Why not use a 5D INR?}
For our problem, we acknowledge that a 5D INR that augments spatial coordinates with the light direction (\eg $(x,y,z,\theta,\phi)$) is a feasible solution, as demonstrated on related tasks in prior work~\cite{bauer2023photon,wu2023hyperinr,chen2025explorable,han2022coordnet}. 
However, such direct INR parameterizations often struggle to cleanly disentangle spatial and conditional features, especially as condition complexity grows. In contrast, diffusion models offer an effective framework for high-dimensional sequence modeling. We therefore investigate diffusion models and view \diffshadow as an initial step toward a pretrained, generative caching paradigm for complex visualization tasks.

\subsection{Stage 1 Training - SIREN Network Overfitting}
To train a diffusion model that compresses and memorizes shadows across lighting directions, we first construct a sufficiently large dataset that captures the distribution of SIREN networks over possible light orientations. We use the Fibonacci-sphere method~\cite{gonzalez2010measurement} to sample lighting uniformly on the sphere.

For each sampled light direction $\boldsymbol{\omega}'$, we independently fit a SIREN network to represent the corresponding shadow coefficient volume. However, in preliminary experiments, we observe that training each SIREN with random initialization leads to unstable behavior in the subsequent diffusion training stage, often resulting in collapsed generations.
We attribute this issue to the non-identifiability of the INR parameterization: the same shadow-coefficient field $g^{\boldsymbol{\omega}'}$ can be represented by many distinct sets of network weights, leading to a highly inconsistent weight space that makes it difficult for the diffusion model to learn a stable distribution.
Similar observations have been reported in prior work~\cite{wu2024blockfusion, wu2023hyperinr}, where the authors mitigate this issue by introducing a shared MLP to enforce a common parameter space across instances. In contrast, our SIREN networks consist solely of fully connected layers without any shared components across different lighting conditions.
To address this limitation, we adopt a preoptimization strategy following HyperDiffusion~\cite{erkocc2023hyperdiffusion}: we first optimize a single SIREN network and use its weights to initialize all other SIREN networks. This significantly reduces weight variation across fitted networks and stabilizes the diffusion training process.

Moreover, jointly optimizing all SIREN networks in a single training run is infeasible due to memory constraints, as training hundreds or thousands of networks simultaneously cannot fit into device memory. To address this, we partition the SIREN networks into subsets and jointly optimize one subset at a time. This strategy enables our method to scale to a large number of SIREN networks representing shadow coefficient volumes under diverse lighting conditions.

\vspace{0.05in}
\noindent
{\bf Loss Functions.} Since the SIREN network is trained to predict shadow coefficient values at arbitrary spatial coordinates, the objective of this stage is to minimize the difference between the network-predicted shadow coefficients and their corresponding ground-truth values. We employ the mean squared error (MSE) loss for this training and sampled 32{,}768 points for a single shadow coefficient volume in each training iteration.

\subsection{Stage 2 Training - Diffusion Model Training}

To model the distribution of shadow fields represented by SIREN networks, we train a diffusion model directly in the space of SIREN networks parameters, which offers several advantages. First, by compressing volumetric data into SIREN networks, the diffusion model can operate on a compact representation, reducing both memory and computational costs compared to conventional 3D generative models. 
Second, this encoder-free representation allows us to treat network parameters as flattened sequential tokens, making them well-suited for transformer-based diffusion architectures. Such architectures can effectively model the complex correlations between weight tokens across different SIREN layers. Prior works~\cite{erkocc2023hyperdiffusion, peebles2022learning} have demonstrated that transformer-based diffusion models can effectively generate MLP weights, further supporting our design choice.

Following the standard denoising diffusion paradigm, Gaussian noise is progressively added to shadow INR weights during training, and a diffusion neural backbone learns to reverse this process. We predict the clean shadow INR weights directly instead of the added noise at each diffusion timestep. During inference, the model samples from a Gaussian prior and iteratively denoises to generate new shadow INR weights consistent with the learned distribution. The forward diffusion process starting from a set of clean shadow INR weights \(z_0\) is defined as:
\begin{equation}
q(z_t \mid z_{t-1}) = \mathcal{N}\left(z_t; \sqrt{1 - \beta_t} \, z_{t-1}, \, \beta_t \mathbf{I}\right)
\end{equation}
where \(z_t\) denotes the shadow INR weights at diffusion timestep \(t\), obtained by adding Gaussian noise to the previous shadow INR weights \(z_{t-1}\). The term \(\sqrt{1 - \beta_t} \, z_{t-1}\) preserves part of the signal from the previous step, and, together with \(\beta_t \mathbf{I}\), they define the mean and variance of the Gaussian distribution \(\mathcal{N}\) used to sample the next shadow INR weights \(z_t\) with additional noise.

\begin{figure}
    \centering
    \includegraphics[width=\linewidth]{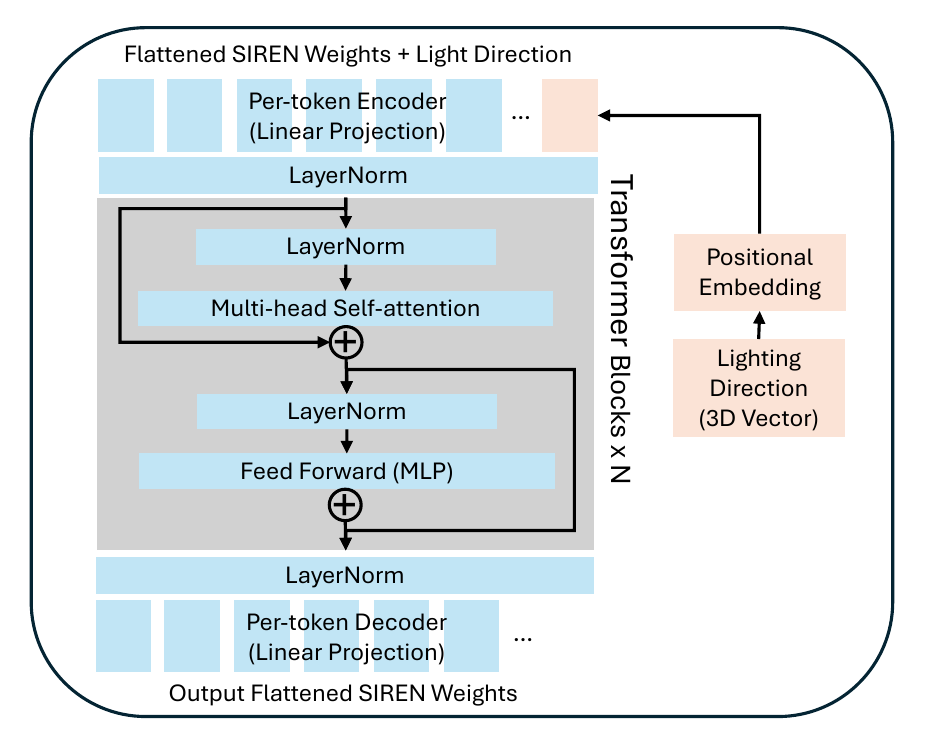}
    \caption{The architecture of our diffusion backbone. A GPT-2-style Transformer conditioned on lighting directions encoded via Fourier positional embedding~\cite{mildenhall2020nerf}. The embedded lighting features are injected into the network as additional input tokens of the transformer (in-context conditioning).}
    \label{fig:diff_arch}
\end{figure}

\vspace{0.05in}
\noindent
{\bf Diffusion Backbone.} 
Our diffusion backbone is a transformer-based architecture (see \autoref{fig:diff_arch}) composed of GPT-style transformer blocks~\cite{vaswani2017attention, radford2019language} without causal masking.
We chunk shadow INR weights layer by layer, where the weights and biases of each layer are treated as individual input tokens. These tokens are mapped to and from the transformer via simple linear projections, allowing bidirectional self-attention to enable cross-layer interactions among SIREN weight tokens.

Once trained, the diffusion model accurately reconstructs shadow INR representations for the requested pre-trained lighting direction by sampling from the learned distribution. This circumvents the need to read and decode thousands of independent networks from disk, highlighting the effectiveness of the diffusion model as a compressed shadow cache.

\vspace{0.05in}
\noindent
{\bf Conditioning.}
To extract the appropriate shadow field, we condition the diffusion model on the lighting direction. Rather than searching through a database of INRs, querying the diffusion model with the pre-trained light vector reconstructs the required shadow INR weights instantly.
The lighting direction is represented as a 3D vector and used as a conditional input. To provide a richer representation, we apply a positional embedding similar to that used in NeRF~\cite{mildenhall2020nerf} to map the lighting vector into a higher-dimensional feature space. These embeddings are then appended to the input SIREN tokens and fed into the transformer (see \autoref{fig:diff_arch}). This type of conditioning is commonly referred to as in-context conditioning in the literature~\cite{dosovitskiy2020image, peebles2023scalable}.
This design enables the network to index its unified weight space and retrieve the accurate shadow coefficients.

\vspace{0.05in}
\noindent
{\bf Basic Loss Functions.} 
As stated above, we let the transformer directly predict the clean shadow INR weights \(z_0\). Hence, the basic training objective is defined as the MSE error between the clean shadow INR weights \(z_0\) and the predicted shadow INR weights \(z_{\theta}\), formulated as follows:
\begin{equation}
L_{\text{diffusion}} = \,\parallel z_0 - z_\theta \parallel ^{2}
\end{equation}

\vspace{0.05in}
\noindent
{\bf Geometry Loss.} 
To further improve shadow quality reconstructed from diffusion model, we introduce a geometry loss that directly supervises the diffusion model in the volumetric domain, rather than solely in SIREN weight space. Specifically, during each diffusion training iteration, we decode the generated shadow INR weights and evaluate them at sampled spatial locations. For each SIREN-volume pair, we pre-sample 500{,}000 spatial points, storing their coordinates and corresponding ground-truth scalar values. This allows efficient supervision without requiring on-the-fly shadow computation or storing full volumetric grids during training.

\vspace{0.05in}
\noindent
{\bf Rendering Loss.} 
While the shadow coefficient volumes decoded from the generated shadow INR weights can exhibit good reconstruction quality, we observe that the corresponding rendered images may still contain noticeable artifacts. This indicates that even small errors in the volumetric domain can be amplified to reflect in the rendered images. Motivated by the success of rendering-based supervision in neural rendering methods such as NeRF~\cite{mildenhall2020nerf}, we incorporate a rendering loss that supervises the model directly in image space by minimizing pixel-wise discrepancies between rendered outputs and ground-truth images.

Specifically, during training, we use the generated shadow coefficients to perform ray marching and render images, leveraging the differentiability of the rendering process to compute the loss and optimize the diffusion model. We pre-render a set of ground-truth images from multiple viewing angles, following a similar idea as NeRF, which learns implicit neural representations from multi-view images. This encourages the diffusion model to generate shadow INR weights that produce consistent renderings across different perspectives of the scene. 

However, evaluating the rendering loss at every training step is computationally expensive in both time and memory. To fit the training process within the available memory while retaining effective rendering supervision, we randomly sample a small subset of pixels (e.g., 24) from each image instead of computing the loss over the full resolution (e.g., 128×128) at every iteration.
Moreover, we adopt a two-stage training strategy by first optimizing the diffusion model with the geometry loss for fast convergence, followed by fine-tuning with the rendering loss to improve the fidelity of the generated shadow INR weights in the rendered domain.

\subsection{Network Design}
In the following, we outline the network design of our pipeline as well as our selection of hyperparameters and considerations for input data processing.

\vspace{0.05in}
\noindent 
{\bf Network Architecture.} 
For the SIREN overfitting stage, we mainly followed the network architecture used in the NeurComp~\cite{lu2021compressive}, whose network also builds on top of the original SIREN work~\cite{sitzmann2020implicit}. 
Our SIREN network consists of 4 fully-connected layers with 128 neurons (64 for smaller datasets) and residual blocks. 
Moreover, the transformer model used in the diffusion training follows the design of HyperDiffusion~\cite{erkocc2023hyperdiffusion}. The number of transformer blocks is 12, the embedding dimension per token is 720, and the number of attention heads is 16.

\vspace{0.05in}
\noindent
{\bf Hyperparameters.} 
All models were trained using the Adam optimizer with \(\beta_{1} = 0.9\) and \(\beta_{2} = 0.999\). The learning rate and its scheduling strategy were tailored to each training stage. For the SIREN network, the initial learning rate was set to \(10^{-3}\) with a cosine decay schedule over 2000 epochs. The diffusion model used a learning rate of \(2 \times 10^{-4}\) with a \texttt{StepLR} scheduler (\(\gamma\) = 0.9, \textit{step} = 200) over 7000 epochs training with geometry loss and additional 2000 epochs training with rendering loss. Lastly, the number of diffusion timestep we used for all datasets is 500.

\vspace{0.05in}
\noindent
{\bf Data Normalization and Standardization.} 
In the SIREN overfitting stage, sampled shadow coefficients already lie in $[0, 1]$, so no additional normalization is needed to improve the training stability and performance. However, we empirically found that performing data standardization on the input shadow INR weights is crucial for stable diffusion training, as it reduces the distribution mismatch between shadow INR weights and the Gaussian noise used in the diffusion process.
\section{Evaluation}
We evaluate our framework using multiple datasets and report both quantitative and qualitative results. The following subsections describe the datasets, system setup, implementation details, and evaluations of visual reconstruction quality, rendering performance, as well as comparisons with baseline approaches.

\begin{figure*}[t!]
\centering
\includegraphics[width=\linewidth]{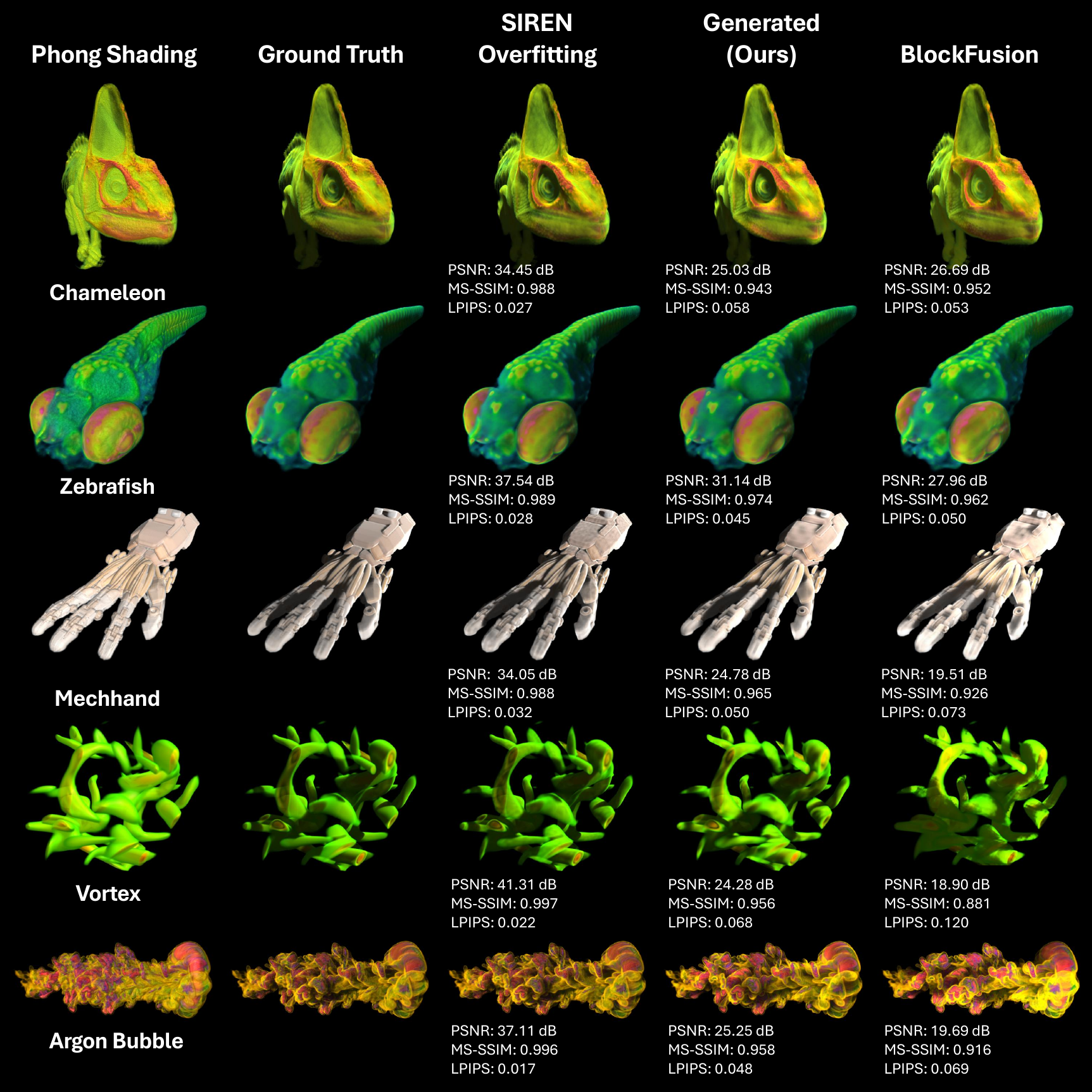}
\vspace{-6mm}
\caption{Visualization of volume renderings for five different datasets (rows) with shadow reconstructed from the SIREN networks across training stages, DVR with Blinn-Phong shading, and baseline method - BlockFusion (columns).}
\label{fig:recon_comparison}
\end{figure*}

\begin{table}[t]
\centering
\caption{Dataset used.}
\small
\label{tab:dataset_resolution}
\resizebox{3.5in}{!}{ 
\begin{tabular}{l c | l c}
\toprule
Dataset & Resolution & Dataset & Resolution \\
\midrule
\textsc{Chameleon} & 1024 $\times$ 1024 $\times$ 1080 &
\textsc{Zebrafish}   & 592 $\times$ 413 $\times$ 956 \\

\textsc{Mechhand}    & 640 $\times$ 220 $\times$ 229 &
\textsc{Vortex}     & 128 $\times$ 128 $\times$ 128 \\

\textsc{Argon Bubble}     & 128 $\times$ 128 $\times$ 256 \\ 
\bottomrule
\end{tabular}
}
\end{table}

\subsection{Datasets}
We evaluated our method on a collection of anatomical, biomedical, and fluid simulation datasets to demonstrate its utility on complex anatomical and fluid structures. Specifically, we include \dataname{Chameleon}, \dataname{Zebrafish}, \dataname{Mechhand}, \dataname{Vortex}, and \dataname{Argon Bubble}, 
which feature both detailed external appearances and intricate internal boundaries.
Several of these datasets particularly benefit from shadow-enhanced visualization, which improves the perception of spatial relationships between anatomical structures compared to direct volume rendering with Blinn-Phong shading. For example, \dataname{Chameleon} features an uneven body surface, where shadow cues enhance depth perception and make important features (e.g., the eyes) more distinguishable. Moreover, shadows in \dataname{Vortex} dataset help reveal the separation between neighboring vortex tubes, where one tube naturally casts shadows onto another.
Some datasets reach resolutions of up to \(1024^3\) voxels, where computing shadow coefficients in real time becomes prohibitively expensive, demonstrating the necessity of a precomputed caching approach. All datasets used in our experiments, along with their respective resolutions, are summarized in \autoref{tab:dataset_resolution}.

\vspace{0.05in}
\noindent
{\bf Training Data Preparation.}
\label{sec:training_data_preparation}
To facilitate diffusion model training, we prepared 2,000 shadow coefficient volumes for each dataset, corresponding to uniformly distributed lighting directions generated using the Fibonacci sphere method. To balance training efficiency and lighting directions' coverage, we restricted the sampled lighting directions to a specific sub-region of the sphere, selecting 2,000 directions from a total of 8,000 points. This sub-region was chosen according to each dataset’s orientation and typical viewing perspective, focusing on the areas most commonly viewed by users (e.g., the front-facing or head region of the dataset).

\subsection{System setup}
All of our experiments ran on the Sophia and Polaris supercomputer at the Argonne Leadership Computing Facility. Sophia comprises 24 NVIDIA DGX A100 nodes, each with eight NVIDIA A100 Tensor Core GPUs providing 40 GB of memory per GPU. Polaris is a 560-node system, where each node is equipped with four NVIDIA A100 GPUs. For SIREN overfitting and diffusion training with geometry loss, each of the datasets only requires a single GPU to run the training. Since the diffusion training finetuned with rendering loss is more computationally expensive in terms of time and memory usage, we scaled it to use 1 node with 4 GPUs on Polaris for each dataset, which allows us to have larger batch size for more stable training and faster convergence.

\subsection{Implementation Details}
\label{sec:impl_details}
Our system is implemented in Python using PyTorch and PyTorch Lightning. We distribute diffusion training across multiple GPUs using the PyTorch Distributed Data Parallel (DDP) module. At inference time, we utilize the Tiny CUDA Neural Networks library~\cite{tiny-cuda-nn} to accelerate SIREN network queries for volume rendering. \rev{To further improve rendering efficiency, we employ macro-cell-based adaptive sampling \cite{novak2014residual, wu2023interactive} along primary rays, reducing unnecessary network evaluations in empty or low-density regions.}

\vspace{0.05in}
\noindent

{\bf Training Performance and Memory Usage.} We report the training time and peak GPU memory usage for each stage of our pipeline. SIREN overfitting time varies with dataset size, with the largest dataset, \dataname{Chameleon}, requiring approximately 1.5 hours, while peak memory usage remains nearly constant at 14.50 GB across datasets because the same number of sample points is used in each training iteration. Diffusion training costs depend on the SIREN network architecture; here, we report results for SIRENs with 128 neurons per layer. Geometry-loss training takes approximately 6.5 hours on a single GPU with a peak memory usage of 19.62 GB, followed by 6.5 hours of rendering-loss fine-tuning on four GPUs, with a peak memory usage of 29.60 GB per GPU.

\subsection{Reconstruction Quality}
We evaluate the reconstruction fidelity of diffusion-produced shadow effects using both quantitative image-based metrics and qualitative visual comparisons. Specifically, we compare rendered images produced using diffusion-generated shadow INRs against those rendered with ground truth shadow coefficients, as presented in \autoref{fig:recon_comparison}. Additionally, we evaluate against several image-based metrics such as Peak Signal-to-Noise Ratio (PSNR), Multiscale Structural Similarity (MS-SSIM) and Learned Perceptual Image Patch Similarity (LPIPS).

To provide a complete picture of reconstruction quality in the full training pipeline, we provide the results from the two training stages: the SIREN networks from the overfitting stage and diffusion training stage. Comparing these sets allows us to analyze differences and how reconstruction quality varies across the training pipeline.

\subsubsection{Visual Comparison}

Since the shadows produced in the SIREN overfitting stage serve as training targets for the diffusion model, we first examine the rendered results from this stage. As shown in \autoref{fig:recon_comparison}, the overfitting stage achieves high-fidelity reconstructions with sharp structural boundaries and well-preserved local details across datasets. We also observe slight grainy artifacts on the flat surfaces of \dataname{Mechhand}; however, the reconstruction still achieves a PSNR of over 34 dB.

For the final results from diffusion training, all datasets largely preserve the overall shadow structure (e.g., large shadow regions), with \dataname{Zebrafish} showing the closest visual match to both the ground truth (31.14 dB PSNR) and the SIREN overfitting results. Nevertheless, some outputs exhibit minor color shifts, and certain fine details appear slightly smoothed. 
For example, the shadows in \dataname{Vortex} appear slightly blurred in regions with fine shadow structures, although the overall shadow distribution is well preserved.

For additional context, we also include a comparison with ray marching-based volume rendering with Blinn-Phong shading, which highlights how incorporating shadow effects significantly enhances the realism of the rendered images. For example, in the \dataname{Chameleon} dataset, the shadow effects emphasize the uneven surface geometry of the head, making the concave regions around the eyes and cheeks much more apparent compared to the no-shadow rendering.

\subsection{Rendering Performance and Runtime Memory Usage}

\rev{A key advantage of our approach is that the predicted shadow INR enables advanced illumination for INR-based volume rendering. To validate this capability, we use a real-time INR-based volume renderer that co-renders the data INR and shadow INR. We evaluate rendering performance in frames per second (FPS) and compare our approach against a na\"ive baseline that computes shadows by performing secondary ray sampling with the data INR, ensuring a fair comparison under our target deployment scenario in which all rendering operations are performed directly on the INR representations rather than the original volume. Both our method and the baseline are optimized with CUDA for network inference and employ macro-cell-based adaptive sampling along primary rays (see \autoref{sec:impl_details}).
The performance results are summarized in \autoref{tab:rendering_perf}. 
The results show our method can achieve significantly faster rendering and lower runtime memory usage across all datasets, demonstrating the benefit of direct querying of precomputed shadow coefficients from the SIREN network. Additional implementation details are provided in the Appendix.}

\begin{table}[htbp]
    \centering
    {\small
    \caption{Rendering performance (FPS) and GPU memory usage comparison between our method and the naive baseline. Higher FPS indicates faster rendering; lower memory indicates a smaller footprint. Our approach demonstrates significant speedups and reduced memory usage across all datasets.}
    \label{tab:rendering_perf}
    \begin{tabular}{lcccc}
        \toprule
        & \multicolumn{2}{c}{\textbf{FPS} $\uparrow$} & \multicolumn{2}{c}{\textbf{Memory (GiB)} $\downarrow$} \\
        \cmidrule(lr){2-3} \cmidrule(lr){4-5}
        \textbf{Dataset} & \textbf{Ours} & \textbf{Na\"ive} & \textbf{Ours} & \textbf{Na\"ive} \\
        \midrule
        Chameleon      & 2.864  & 0.021 & 8.788 & 18.664 \\
        Zebrafish      & 11.125 & 0.139 & 5.342 & 15.282 \\
        Mechhand       & 11.510 & 0.284 & 3.476 & 9.972  \\
        Argonne Bubble & 12.155 & 0.511 & 3.748 & 11.188 \\
        Vortex         & 21.815 & 0.616 & 3.108 & 8.656  \\
        \bottomrule
    \end{tabular}
    }
\end{table}

\subsection{Baseline Comparison}
To assess the reconstruction quality against existing methods, we choose BlockFusion~\cite{wu2024blockfusion} and Deep Volumetric Ambient Occlusion~\cite{engel2020deep} as the baseline, as both address 3D content prediction.

\begin{figure}
    \centering
    \includegraphics[width=\columnwidth]{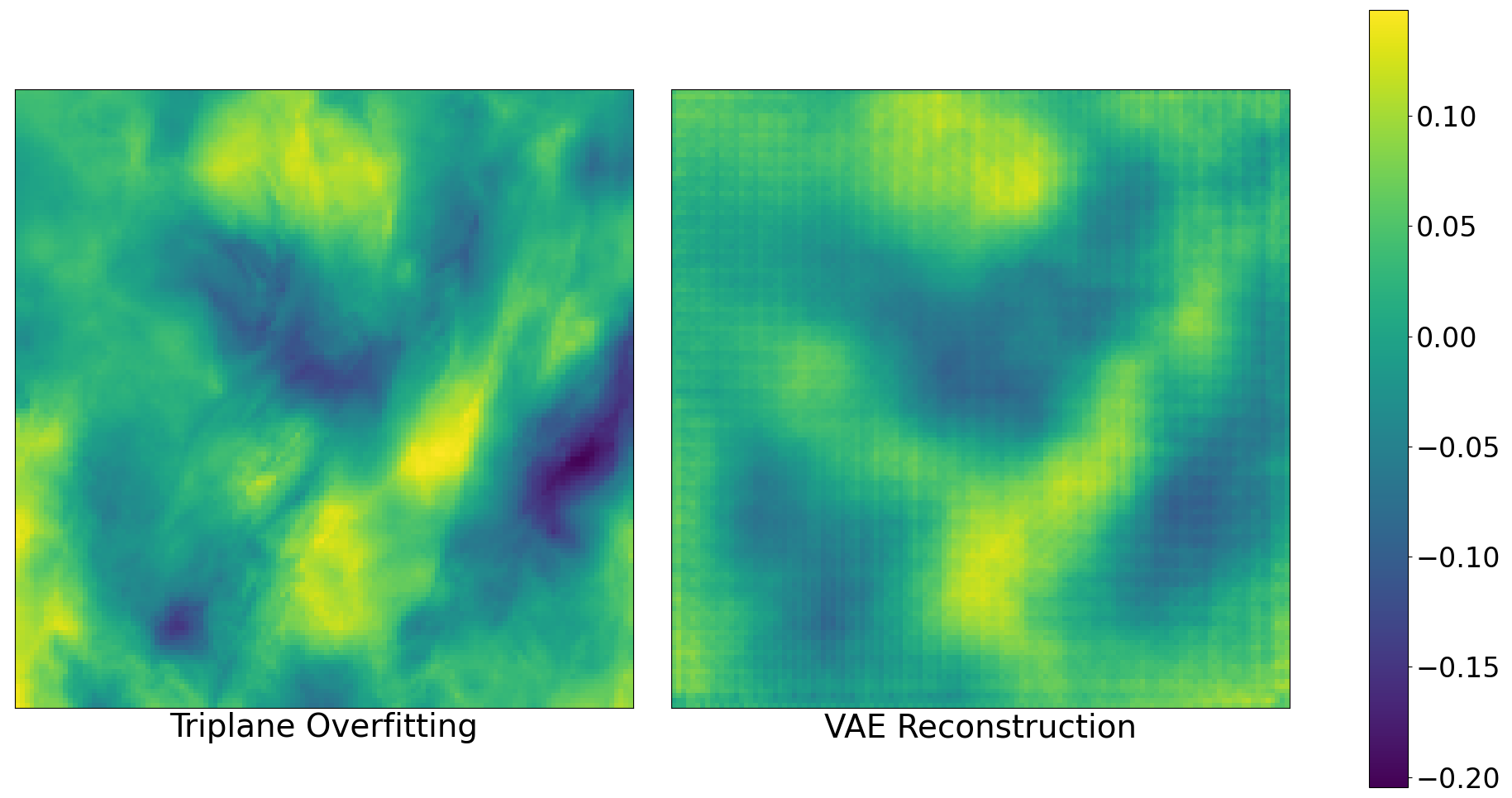}
    \vspace{-1.5em}
    \caption{
    Failure Case of VAE-Reconstructed Triplanes. Visualization of \dataname{Vortex} triplanes from the triplane overfitting stage (left) and the VAE reconstruction stage (right). The overfitted triplanes are used as training targets for the VAE. However, the VAE fails to preserve high-frequency structural details.
    }
    \label{fig:blockfusion_triplane_compare}
\end{figure}

\subsubsection{BlockFusion}
BlockFusion is a 3D scene generation method based on a latent diffusion model, where triplane encoding~\cite{chen2022tensorf} is used to represent 3D geometry, and a VAE projects this representation into a latent space for diffusion training. This representation makes BlockFusion a suitable baseline for our work, as diffusion models can be trained to generate triplane-based INRs that represent shadow effects in our application. To adapt BlockFusion to our problem, we use triplane encoding to represent shadow coefficient volumes instead of mesh geometry, and train the diffusion model to predict the corresponding triplane-based shadow INRs. We keep the same diffusion architecture and training procedure as BlockFusion for a fair comparison, but revise the conditioning mechanism to incorporate embedded lighting direction vectors. In contrast to our encoder-free SIREN approach, BlockFusion requires an intermediate VAE compression step. Visualization results using shadow INRs generated by the BlockFusion diffusion model are shown in \autoref{fig:recon_comparison}.

For the datasets such as \dataname{Chameleon} and \dataname{Zebrafish}, BlockFusion produces visually good results, while some shadow regions appear less dark than in the ground truth. 
For \dataname{Chameleon}, it achieves comparable evaluation metrics to our approach.
However, BlockFusion struggles to reconstruct accurate shadows for datasets with more intricate shadow structures, such as \dataname{Vortex}, where even the primary shadow regions are not faithfully recovered (see \autoref{fig:recon_comparison}). 
We attribute this degradation primarily to the VAE stage, which compresses the raw triplanes into latent representations. As shown in \autoref{fig:blockfusion_triplane_compare}, the reconstructed triplanes of \dataname{Vortex} lose high-frequency structural information, and these errors propagate to the subsequent diffusion stage, degrading the final reconstruction quality.

Beyond reconstruction quality, our approach also offers substantially lower training cost than BlockFusion. Among BlockFusion's three training stages, the VAE stage alone requires 60--75 hours on 8 GPUs, far longer than the roughly 5-hour triplane overfitting stage and 14-hour diffusion training stage on a single GPU. Despite this computational cost, it still fails to consistently produce high-quality reconstructions across all datasets. 

In contrast, our full training pipeline completes in less than one day using at most 4 GPUs, demonstrating substantially higher training efficiency than BlockFusion.

\begin{figure*}
    \centering
    \includegraphics[width=\textwidth]{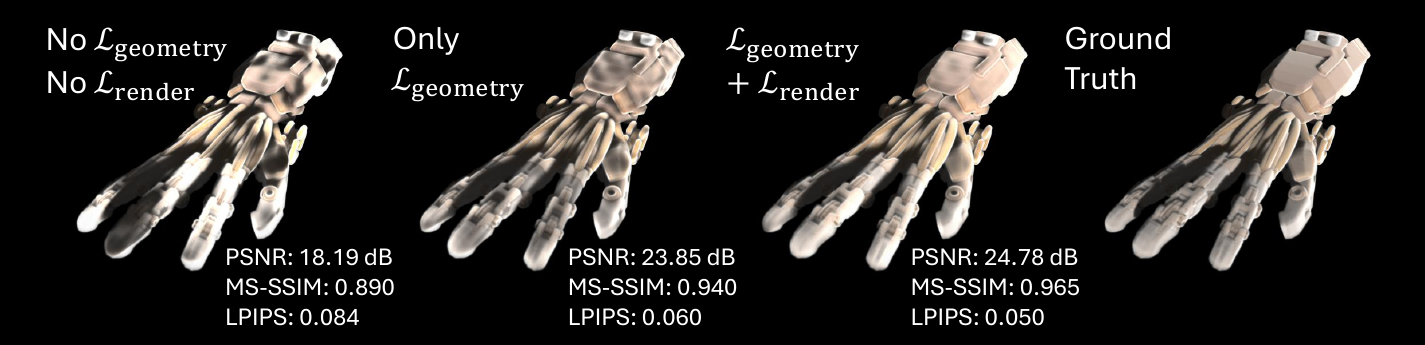}
    \vspace{-1.5em}
    \caption{Diffusion-reconstructed shadows for \dataname{Mechhand} trained with different auxiliary loss configurations: no geometry/rendering loss (left), geometry loss only (center-left), geometry + rendering loss (center-right), and ground truth (right). From the first three sub-figures, we observe progressive visual improvement as auxiliary losses are added, with spotty artifacts gradually reduced. This result suggests the importance of incorporating volume-domain-supervision and rendering supervision in diffusion training.}
    \label{fig:ablation_compare}
\end{figure*}

\subsubsection{Deep Volumetric Ambient Occlusion}
Deep Volumetric Ambient Occlusion (DVAO) is a related work that uses deep learning techniques to predict precomputed illumination volumes—specifically ambient occlusion volumes, whose values depend on the transfer function—and are later used in direct volume rendering (DVR). Since our work also focuses on precomputed illumination volume prediction, DVAO serves as a relevant baseline for comparison. Unlike our approach, which operates on INR weights, DVAO utilizes a 3D U-Net architecture~\cite{ronneberger2015u, cciccek20163d} that directly takes raw scalar volumes as input and outputs the corresponding ambient occlusion volumes given a specific transfer function information.

To evaluate how well this volume-to-volume prediction framework generalizes to our problem, we adapt their approach to predict shadow coefficient volumes. DVAO proposes several conditioning strategies to incorporate transfer function information into the U-Net. Among these, we adopt the Latent Concatenation strategy, which injects encoded conditioning features into the latent (bottleneck) layer of the network. This strategy is the most suitable for our setting, as it allows us to incorporate light direction information in a similar manner by encoding it with positional embeddings and injecting it into the latent space.

We set up experiments following their approach by training on 2,000 shadow instances for each dataset, \dataname{Zebrafish} and \dataname{MechHand}, separately. Different from their original training setting, where multiple volumes are used as input, our task focuses on a single dataset. Therefore, the same raw volume from either \dataname{Zebrafish} or \dataname{MechHand} is used as the input to the U-Net, while different light directions are injected as conditioning variables and the corresponding shadow coefficient volumes are used as training targets. After training, we observe a consistent behavior across both datasets: the network produces nearly identical outputs despite different lighting directions being provided as conditioning input. This suggests that latent-space conditioning has limited influence on the U-Net output in this setting.

We note that this comparison is not entirely equivalent to the original DVAO setting. In DVAO, the network is trained on multiple input volumes of the same data type (e.g., CT head scans), whereas in our problem each dataset has significantly different geometric structures, and the model is trained per dataset. As a result, the network receives the same volume as input and must rely primarily on the conditioning variable (light direction) to produce different outputs. This one-to-many mapping is difficult for a deterministic 3D CNN to learn, which may lead the network to converging to an average solution that ignores the conditioning signal. In contrast, diffusion models learn a conditional distribution instead of a deterministic mapping, which enables the generation of diverse outputs from random Gaussian noise given different conditioning inputs.

Moreover, training a 3D CNN is highly memory-intensive. To fit within GPU memory constraints, the input volumes often need to be downsampled to lower resolutions (e.g., \(64^3\) or \(128^3\)) for training, which inevitably sacrifices reconstruction quality and fine structural details. In contrast, our approach encodes volumes into INRs, where diffusion models are applied. This significantly reduces memory requirements and enables our method to scale to higher-resolution volumes while maintaining the reconstruction quality.



\section{Ablation Study}
In this ablation study, we evaluate the impact of key design choices in our training pipeline, including the geometry loss, rendering loss, and the sampling density of lighting directions, on shadow reconstruction quality.

\subsection{Diffusion with Geometry Loss}
We first ablate the geometry loss from the diffusion model training and compare the resulting shadow effects with those produced when geometry loss is included. As shown in the 
left-most image in  \autoref{fig:ablation_compare}, the results trained without auxiliary losses, particularly geometry loss, show significant artifacts and noticeable color shifts in the reconstructed shadows. This observation further demonstrates the importance of volume-domain supervision in guiding the diffusion model.

\subsection{Diffusion with Rendering Loss}
Furthermore, we conduct an ablation study by removing the rendering loss from diffusion training and comparing the results with those obtained when both geometry and rendering losses are used. As shown in \autoref{fig:ablation_compare}, the spotty artifacts observed on \dataname{Mechhand} are reduced when rendering loss is included. This result indicates that incorporating pixel-wise rendering loss helps improve the visual quality of the rendered images with reconstructed shadows.

\subsection{Effect of Lighting Direction Sampling Density}
\label{sec:ablation_study_light_dirs_density}

Finally, we investigate how the sampling density of lighting directions affects the reconstruction quality of shadow INRs produced by the diffusion model. 
We train two models on \dataname{Zebrafish}, one using 2,000 and the other 6,300 sampled lighting directions, with both sets sampled from the same spatial region.
To efficiently evaluate a large number of shadow instances, we compute voxel-wise PSNR from sampled points in the reconstructed shadow coefficient volumes and report the average over each group.

We then evaluate each model under two testing conditions: (1) lighting directions seen during training, and (2) an additional set of 500 unseen lighting directions sampled from the same region. This setup allows us to assess both reconstruction performance on observed lighting conditions and the model’s ability to predict to unseen directions within the same lighting space. The results are summarized in \autoref{tab:light_direction_density_exp}. 

The first observation is that the average reconstruction quality on the training set is very similar when training with 2,000 and 6,300 lighting directions, with PSNR values of 29.57 dB and 29.32 dB, respectively. This suggests the potential of our approach to scale to a larger number of lighting conditions with minimal reconstruction quality degradation, which highlights the strong memorizing capacity of the diffusion model. However, the performance of the model on unseen test lighting directions is consistently lower than that on the training set. Our further analysis shows that reconstruction quality degrades as the unseen lighting directions become farther from the training lighting directions in the lighting space. We believe that more advanced conditioning strategies for diffusion models could help address these limitations and improve the quality to those unseen lighting conditions.

\begin{table}[htbp]
    \centering
    {\small
    \caption{Average reconstruction quality (PSNR) of shadow coefficient volumes for the training and test sets using 2,000 and 6,300 lighting instances. Both training sets are sampled from the same spatial lighting region. An additional 500 lighting directions sampled from the same region are used for the test set.}
    \label{tab:light_direction_density_exp}
    \begin{tabular}{ccccc}
        \toprule
        \textbf{\shortstack{Number of Lights\\in Train Set}} & \textbf{Train Set Lights} & \textbf{Test Set} \\ 
        \midrule
        2000 & 29.57 & 20.37 \\
        6300 & 29.32 & 20.70 \\
        \bottomrule
    \end{tabular}
    }
\end{table}
\section{Discussion}
While our diffusion-based shadow caching framework demonstrates strong performance across multiple datasets, our experiments also reveal several limitations that suggest opportunities for 
improvement.

\begin{figure}
    \centering
    \includegraphics[width=\columnwidth]{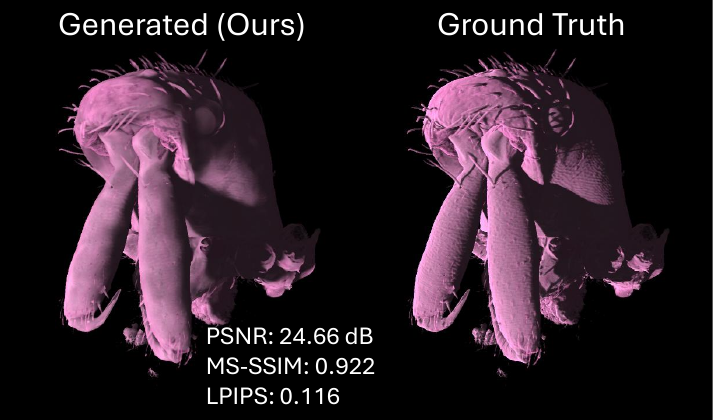}
    \vspace{-1.5em}
    \caption{\rev{Limitations on high-frequency shadow reconstruction. While the overall illumination of \dataname{Spider} dataset remains plausible, fine shadow structures are over-smoothed, illustrating the difficulty of representing high-frequency shadow discontinuities with deep neural networks.}}
    \label{fig:spider_limitation}
\end{figure}

The primary factor limiting the fidelity of the diffusion-reconstructed shadows is the smoothing effect inherent in learning a vast, continuous weight space. 
Although our encoder-free method successfully bypasses the blurring introduced by the VAE in the baseline, directly predicting raw shadow INR weights with a Transformer can occasionally over-smooth high-frequency shadow details, as illustrated in \autoref{fig:spider_limitation}.
This occurs because the diffusion model learns a smoothed distribution over the parameter space, which sometimes suppresses the sharp structural boundaries captured by individual overfitted SIREN networks.

A natural direction to address this limitation is to increase the capacity of both the SIREN and diffusion models to better memorize high-frequency variations. Additionally, more expressive, multi-scale geometry-aware losses for diffusion training could further preserve sharp volumetric structures, although their high computational cost remains a challenge. Developing efficient sampling strategies for these losses may offer a practical balance between reconstruction fidelity and training efficiency.

More broadly, diffusion models offer the potential for a unified caching framework that compresses multiple varied volumetric datasets into a single model without per-dataset training. We also envision extending the framework to other precomputed illumination fields, such as ambient occlusion, to demonstrate the broader applicability of diffusion-based shadow caching.

\section{Conclusion}
We introduce \diffshadow, a novel shadow caching framework for shadow coefficient volumes in direct volume rendering. Unlike conventional approaches that require costly ray-marching evaluations or demand immense storage for precomputed grids, \diffshadow uses a conditional diffusion model to cache thousands of lighting-dependent shadow fields into a unified weight space. By querying the diffusion model, our method instantly reconstructs the required shadow INR weights, acting as a highly compressed memory cache. The reconstructed shadow INR can then be co-rendered with the data INR, replacing expensive secondary-ray evaluation and substantially accelerating INR rendering with realistic shadows.
Our findings highlight the effectiveness of diffusion models as a dense shadow cache and suggest future directions for scalable, memory-efficient illumination modeling in volume rendering. As neural rendering transitions from a research concept to a foundational technology, our work offers a practical approach for managing complex, multi-dimensional volumetric assets in demanding scientific applications.
\section*{Acknowledgments}
This work was supported by the U.S. Department of Energy, Office of Science (SC), Advanced Scientific Computing Research (ASCR), Competitive Portfolios Project, PIONEER:  Program for Intelligent Optimization for Next-generation Experiments, Exploration, and Research, under Contract DE-AC02-06CH11357. This research used resources of the Argonne Leadership Computing Facility, which is a U.S. Department of Energy Office of Science User Facility operated under contract DE-AC02-06CH11357.

\bibliographystyle{elsarticle-num}
\bibliography{references}

\appendix
\section{Rendering Pipeline and Baseline Shadow Implementation}
\label{app1}

Our INR-based volume renderer follows a wavefront-style \cite{laine2013megakernels} rendering architecture (ray marcher), similar to \cite{wu2023interactive}, composed of several specialized kernels: a coordinate kernel that collects sample-point coordinates along rays, a network inference kernel that batches these points through the neural network, and a shading kernel that performs color composition. The ray marcher processes ray samples in small batches, running all three kernels in sequence for each batch, and repeats until a per-ray termination criterion is satisfied. This design is well suited to machine learning frameworks, where network inference is naturally batch-oriented, and it avoids out-of-memory issues that would arise from inferring all sample points along all rays at once.

To incorporate shadow effects via an INR, we additionally infer a shadow INR alongside the primary network inference step to obtain shadow coefficients, which are then combined with the scalar field values to modulate shading during composition.

For our baseline method, we introduce an additional, nested instance of this batched marching pipeline to evaluate shadow coefficients for each primary-ray sample point via secondary-ray marching. Specifically, after collecting a batch of primary-ray sample points, we pause primary-ray marching and diverge into an inner loop that iteratively marches the corresponding secondary rays in batches until they terminate. Only once shadow coefficients have been obtained for every sample point in the current primary batch does marching resume for the next batch of primary rays. This nested-loop structure, combined with the added cost of network inference for the secondary rays, makes the baseline's rendering performance substantially lower, as shown in \autoref{tab:rendering_perf}.

\end{document}